\documentclass[11pt]{article}
\usepackage[T1]{fontenc}
\usepackage{tgbonum}
\usepackage{amsmath,slashed,graphicx,xcolor,bbm,hyperref}
\usepackage[leftcaption]{sidecap}
\DeclareMathAlphabet{\boldmathe}{T1}{cmr}{bx}{it}
\newcommand{\mbf}[1]{\boldmathe{#1}}
\newcommand{\vb}{\mbf{b}}
\newcommand{\vh}{\mbf{h}}
\newcommand{\vj}{\mbf{j}}
\newcommand{\vA}{\mbf{A}}
\newcommand{\vB}{\mbf{B}}
\newcommand{\vD}{\mbf{D}}
\newcommand{\vJ}{\mbf{J}}
\sidecaptionvpos{figure}{c}
\def\mtxt#1{\quad\text{#1}\quad}
\begin{document}
	\begin{titlepage}
		\title {
			\hfill{\small preprint DIAS-STP-92-46}\\[15mm]
	%		\hfill{\small preprint DIAS-STP-85-21}\\[15mm]
			Aharonov-Bohm effect in Presence of Superconductors}
		\author{
			L. O'Raifeartaigh\\
			\emph{\small Dublin Institute for Advanced Studies, 10 Burlington
			Road, Dublin 4, Ireland}\\[3mm]
			 N. Straumann\\
			\emph{\small University of Z\"urich, Sch\"onberggasse 9, CH-8001 Z\"urich, Switzerland}\\[3mm]
			A. Wipf\footnote{present address:
			Theoretisch-Physikalisches-Institut, Friedrich-Schiller-Universit\"at Jena, 07743 Jena,
			Germany; email:
			wipf@tpi.uni-jena.de}
			%; url:
			%\url{http://www.tpi.uni-jena.de/qfphysics/homepage/wipf/index.html}
			\\[1mm]
		\emph{\small Institut f\"ur Theoretische Physik,
			Eidgen\"ossische Technische Hochschule}\\
		\emph{\small H\"onggerberg, Z\"urich CH-8093, Switzerland}
	}
 \date{\small February 1993; Latex-ed: 1. May  2023}
		\maketitle
		\begin{abstract}
			\noindent
			The analysis of a previous paper, in which it
			was shown that the energy for the Aharonov-Bohm effect could be
			traced to the interaction energy between the magnetic field of the
			electron and the background magnetic field, is extended to cover the
			case in which the magnetic field of the electron is shielded from
			the background magnetic field by superconducting material. The
			paradox that arises from the fact that such a shielding would
			apparently preclude the possibility of an interaction energy is
			resolved and, within the limits of the ideal situation considered,
			the observed experimental result is derived.

		\end{abstract}
		\vskip5mm
		\begin{center}\small
%			Keywords: vacuum decay; functional determinant; tunneling;
%			instanton; bounce;\\ semiclassical expansion  \\[8mm]
			Published version: Foundations of Physics, Vol. 23
			(1993) 703-709\\
			 doi: 10.1007/BF01883804\\
			arXiv-ed February 18, 1993 ; \LaTeX-ed May 1, 2023
%			No. 5, 1986, p. 703\\%4-44\\[1mm]
%			doi: 10.1016/0550-3213(86)90363-9\\[1mm]
%			arXiv-ed September 1985 ; \LaTeX-ed January 10, 2022
		\end{center}
		Uploaded to arXiv to help researchers with limited library facilities
	\end{titlepage}

	\tableofcontents

\section{Introduction}\label{sec:introduction}

In a recent paper \cite{1} some of the traditional
mystique surrounding the Aharonov-Bohm effect was removed by the
observation that the
energy $\int  \vA\cdot\vj_e$ which produces the effect could be traced to the
interaction energy $\int \vB\cdot\vb_e$ between the background magnetic field
$\vB$ and the magnetic field $\vb_e$ of the electron. The puzzle
posed by the fact that the effect takes place in spite of the fact
that the electron does not interact locally
with $\vB$ is then explained by the fact that the \emph{magnetic field} of the
electron does interact locally with $\vB$ and the question as to whether
the gauge-potential is necessary for a description of the A-B effect is
answered by the statement that it is necessary only
if one requires a local description in terms of the electron.
A more recent article \cite{2} comes to similar conclusions.

As pointed out by Tassie \cite{3}, however, the  $\int \vB\cdot\vb_e$
explanation would appear to encounter some difficulty
in the case that $\vb_e$ is shielded from $\vB$ by
a superconductor, since in that case the integral $\int \vB\cdot\vb_e$
is zero but there is reliable  experimental evidence \cite{4}
to show that the A-B effect occurs. Indeed, due to the double-charge
of the Cooper pairs, the effect occurs in this case
 with maximal phase-shift $\pi$. The difficulty also manifests
itself in the question as to how $\int  \vA\cdot\vj_e$ can be
zero while $\int \vB\cdot\vb_e$ is not zero. The purpose
of the present paper is to clarify this apparent paradox.

\section{Recall of Original Argument}\label{sec:recall}

\begin{figure}[h!]
\begin{center}
\includegraphics{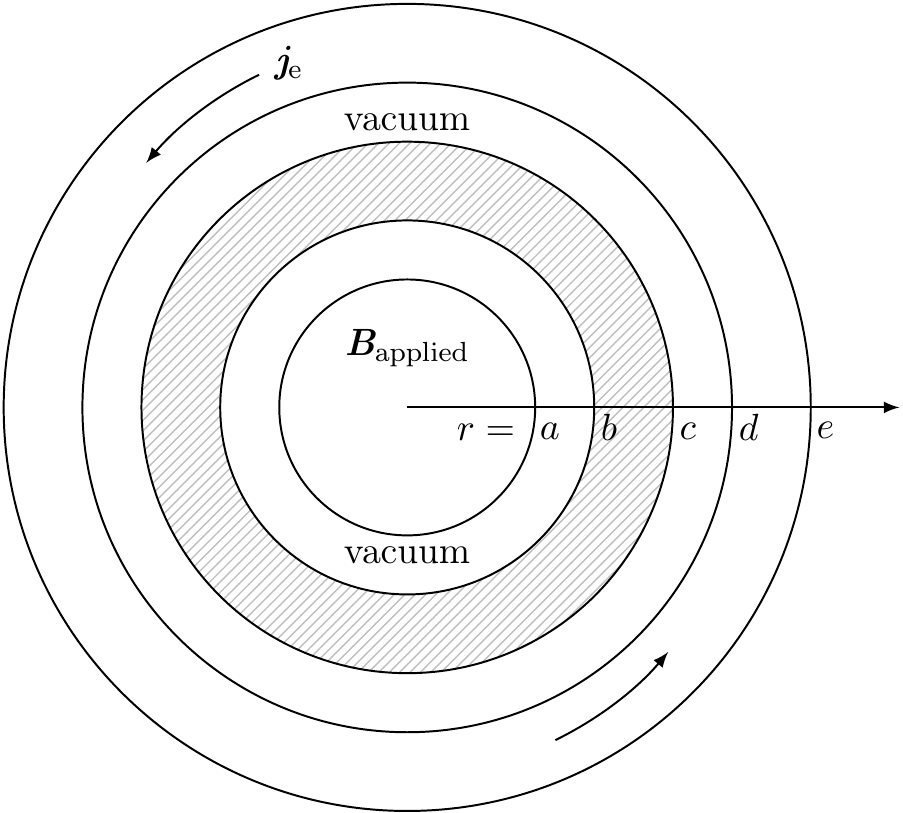}
\caption{\emph{An electron in $d\leq r\leq e$ is circling around
	a magnetic field confined in a long cylinder $r\leq a$.
	The superconductor is in the region $b\leq r\leq c$. In the
	original case the superconductor is removed and there is
	vacuum in this region.}}\label{solenoid}
\end{center}
 \end{figure}

We first give a brief recall of the argument for the
non-superconducting case. Stripped to its essentials the situation
may be described by the diagram in Fig. \ref{solenoid} in which 
the $\vB$-field is
trapped within the cylinder $r= a$ and the electron between the
concentric cylinders $r= d$ and $r= e$. Here everything is
assumed to be uniform in the $z$-direction (direction of the common
axis) and axially symmetric and for the moment there is nothing
between the cylinders $r= a$ and $r= d$, in particular nothing (for
the moment) between
the cylinders $r= b$ and $r= c$. The reason for confining the electron
 by the outer cylinder $r= e$ instead of just leaving it outside
$r=d$ is that, following refs. \cite{1,5}, we
can then consider the AB-Effect as an energy shift computed from the
static Schr{\"o}dinger equation
\begin{equation}
\Bigl(-{d^2 \over dr^2}-\frac{1}{r}\frac{d}{dr} +
{(l+F)^2 \over r^2}\Bigr)\psi(x)  =2mE\psi(x)
\qquad   \psi(d)=\psi(e)=0\,,\label{1}
\end{equation}
rather than a phase shift. Here we suppressed the trivial dependence
on $z$ and $F$ denotes the magnetic flux measured in units of
the elementary flux quantum. The parameter $l$ is an integer
because, as discussed in refs. \cite{1,5}, the electron wave-function
$\psi(x)=e^{il\varphi}\psi(r)$ is single-valued. According to \eqref{1}
 a measurable energy shift is produced
 for non-integral $F$ and this is the analogue of the usual A-B
phase-shift. The potential
$(l+F)^2$ in \eqref{1} which produces this energy shift
derives from an interaction energy of the form
\begin{equation}
\int  \vA\cdot\vj_e \mtxt{where} \vB=\nabla\times \vA \mtxt{and}
\vj_e=\frac{e}{2mi}\big(\bar \psi \vD\psi
-cc\big). \label{2}
\end{equation}
Here we have set $\vb_e=\vh_e$ since no material is
present besides a possible coil.
The point of our earlier paper was that, by considering the integral
in \eqref{2} as over the whole 2-plane, one could use partial integration to
convert it to the form
\begin{equation}
\int \vB\cdot\vb_e \mtxt{where}\nabla\times \vb_e=4\pi\vj_e\,.
\label{3}
\end{equation}
In this formulation the gauge potential is eliminated and
the energy which produces the energy-shift in the
Schr{\"o}dinger equation is traced to the interaction energy between the
two magnetic fields.

\section{The Superconducting Case}\label{sec:superconducting}

To analyze the superconducting case we consider the same idealized situation
but with the intermediate region $b\!\leq\! r \!\leq\! c$ filled with
superconducting material, which would shield the magnetic field
$\vb_e$ of the electron from the background field $\vB$.
Here (in contrast to the situation envisaged in a remark made at the
conclusion of ref. \cite{1} it is assumed that the cooling of the superconductor
has been carried out \emph{after} the field $\vB$ has been switched on so
that $\vB$ is not expelled \cite{6}. In this situation $\vB\!\not\!=0$
but since the magnetic field should be zero in the superconducting region
$b\!\leq \!r \!\leq\! c$ it is difficult to see how the fields $\vB$
and $\vb_e$ could overlap and hence how the integral $\int
\vB\cdot \vb_e$ could be non-zero. This is the paradox mentioned in the
introduction.

To investigate the paradox we have to  consider the field equation
for the $\vB$ field in more detail. This field equation has been given
in \cite{7} and may be written as
\begin{subequations}
\begin{equation}
\frac{d}{dr}\Bigl(\frac{1}{r}\frac{d}{dr}A_{\varphi}\Bigr)=4\pi j_{\varphi}
+4\pi j_e\qquad (\hbox{from}\quad\nabla\times\nabla\times \vA=4\pi \vJ)
\label{4a}
\end{equation}
where
\begin{equation}
j_{\varphi}(r)=\frac{\beta^2}{r}
\Bigl(A_{\varphi}-l\Big(\frac{\hbar}{2e}\Big)\Bigr)\;,
\qquad\beta^2=\frac{4\pi\rho e^2}{mc^2},  \label{4b}
\end{equation}
\end{subequations}
and $\rho$ is the density of the Cooper pairs. Here $A_\varphi$ denotes
the dimensionless $\varphi$-component of the gauge potential so that
the magnetic flux inside radius $R$ is just
\begin{equation}
\Phi_R\equiv\int\limits_{r\leq R}B_z rdrd\varphi=\oint A_\varphi d\varphi.
\label{5}
\end{equation}
As shown in \cite{7} the Ampere equation in the London limit 
(4) can be solved
exactly in terms of Bessel functions. But in the case that the penetration
depth $\beta^{-1}$ is small compared to the width of the superconductor
(which we are assuming to be the case) a better insight into the solution
is obtained if we approximate the Bessel functions by exponentials
and thus approximate the solution by
\begin{align}
2\pi A_\varphi &=\Phi_a\,r^2/a^2&\quad r\leq  a\nonumber \\
&=\Phi_a-\alpha\beta^2(r^2-a^2)(\Phi_a-\Phi_q)&\quad a\leq r\leq b\nonumber \\
&=\Phi_q +2\alpha\beta\sqrt{br}\,(\Phi_a\!-\!\Phi_q)
e^{-\beta(r-b)}+{2\pi\over \beta}\sqrt{cr}\,b_e\,e^{\beta(r-c)}
&\quad b\leq r \leq c\nonumber \\
&=\Phi_q+\pi b_e(r^2-c^2+2c/\beta)&\quad c\leq r\leq d\nonumber \\
&=\Phi_q+\pi b_e(r^2_e-c^2+2c/\beta)&\quad r\geq e.
\label{6}
\end{align}
where $\alpha=(2\beta b\!+\!(\beta b)^2\!-\!(\beta a)^2)^{-1}$. The
flux
\begin{equation}
\Phi_q={l \over 2}\Big({hc\over e}\Big)\label{7}
\end{equation}
is a half-integer multiple of the elementary flux quantum.
Here $l$ is the integer that minimizes $(\Phi_q\!-\!\Phi_a)$, and $b_e$
the $z$-component of the magnetic field due to the electron. The radius
$r_e$ in the gauge potential on the outer region belongs to the
circling electron. All feedback effects are
neglected and the factor one-half in $\Phi_q$ comes from the fact
that the Cooper pairs have charge $2e$.

\section{The Paradox}\label{sec:paradox}

From equations \eqref{6} and \eqref{7} we see that:
\begin{itemize}
\item[a)] Once the superconductor is deeply
penetrated the flux becomes quantized to $\Phi_q$.
\item[b)] There are two
surface currents $\sim\alpha(\Phi_a\!-\!\Phi_q)e^{-\beta(r-b)}$ and
$\sim b_e/\beta\;e^{\beta(r-c)}$.
\item[c)] Whereas the outer surface current vanishes
when the electron current sustaining it is removed ($\vb_e= 0$)
the inner current remains even when the background field is switched
off ($\vB= 0$).
\item[d)] The magnetic field is zero deep within the superconductor.
\item[e)] There is a maximal A-B effect for odd $l$, in agreement with
experiment.
\end{itemize} 
\noindent So the question is: what is the source of the energy for this
maximal A-B effect?

\section{Toy Model}\label{sec:toy}

In order to solve the paradox it is instructive to
remove  the superconducting material from the (shaded)
region $b\!\leq\! r\!\leq \!c$
for a moment and replace it by a particle circling like the
electron, but with charge $q$ say (or more accurately, a $z$-independent
homogeneous charge distribution). Then we have the conventional
A-B effect described in section \ref{sec:recall} for \emph{each} of the
two particles and the source of the energy for the effect
in the respective cases is
\begin{equation}
\int  \vA\cdot\vj_e =\int \vB\cdot\vb_e  \quad\text{and}\quad \int
\vA\cdot \vj_q=\int \vB\cdot\vb_q \,.\label{8}
\end{equation}
The total energy which produces these effects is therefore
\begin{equation}
\int \vB\cdot \big(\vb_e+\vb_q\big)\,.\label{9}
\end{equation}
Suppose now that the particles are oppositely charged and their
angular momenta are such that $\vb_e+\vb_q= 0$. Then the
total energy is zero. Nevertheless there is an A-B effect for each
particle, the energy shift being positive for one and negative for
the other.\par
One arrives at the same conclusion if one puts some diamagnetic
material in the region $b\!\leq\! r\!\leq\! c$. Then 
\eqref{8} is
replaced by
\begin{equation}
\int \vA\cdot\vj_e=\int\vB\cdot\vh_e\,.\label{10}
\end{equation}
The point is that contrary to $\vb_e$ the field $\vh_e$ penetrates
the diamagnet and even in the limit of a perfect diamagnet, $\mu\to 0$,
has nonzero overlap with the external magnetic field.

\section{Resolution of the Paradox}\label{sec:resolution}

Now let us return to the superconductor. From the point of view
of the toy model we see that the superconductor is nothing but a
device to ensure the existence of two oppositely charged currents
whose magnetic fields cancel in the interior regions.
The only difference is that whereas the electron current is the
usual one the q-current is replaced by the
outer surface-current of the superconductor.
If the fields produced by these two currents are denoted by $\vb_e$
and $\vb_s$ respectively, we have, as in the toy model,
\begin{equation}
\vb=\vb_e+\vb_s=0 \mtxt{for} b\leq r \leq c\,.\label{11}
\end{equation}
Thus the total magnetic field inside the superconductor is zero,
as it should be, but the individual fields $\vb_e$ and $\vb_s$  are not.
But the individual magnetic fields $\vb_e$ and $\vb_s$ penetrate the
superconductor and produce the A-B energy by
interacting with $\vB$. It is only the \emph{sum} of these two fields,
which is the total magnetic field, that is zero inside.

Inside the superconductor we have, of course, the situation that
(a) the fields  $\vb_e$ and $\vb_s$ are only virtual, since it is only
their sum that can be measured experimentally
and (b) that the gauge-field is necessary because it is the only
means by which information is transmitted across the superconductor.
Both of these points are true, and the second is an argument in
favour of the necessity of the gauge-potential in the
superconducting case. It is worth observing however
that (a) the fields $\vb_e$ and $\vb_s$ are distinct in the sense that
$\vb_e$ can be measured separately in the region  $c\!<r\!<d$ just
outside the superconductor and (b) within the superconductor
the magnetic potential is no longer
a true gauge-potential because, through the Higg's mechanism, it
has become massive. From this point of view the gauge
potential becomes necessary only when it ceases to be
 a true gauge-potential!

\end{document}